\documentclass[aps,pre,twocolumn,twoside, amsmath,amssymb,aps,bbm,floatfix,showpacs]{revtex4-1}

\usepackage{xcolor}
\usepackage{graphicx}
\usepackage{subfigure}
\usepackage{dcolumn}
\usepackage{bm}
\usepackage{framed}
\usepackage{hyperref}
\usepackage{dsfont}
\usepackage[normalem]{ulem}
\usepackage{nicefrac}

\usepackage{nicefrac}
\usepackage{braket}

\newcommand{\be}{\begin{equation}}
\newcommand{\ee}{\end{equation}}
\newcommand{\bea}{\begin{eqnarray}}
\newcommand{\eea}{\end{eqnarray}}
\newcommand{\la}{\left\langle}
\newcommand{\ra}{\right\rangle}

\begin{document}

\title{Heat distribution of a quantum harmonic oscillator}
\author{Tobias Denzler}
\author{Eric Lutz}
 \affiliation{Institute for Theoretical Physics I, University of Stuttgart, D-70550 Stuttgart, Germany}

\begin{abstract}
We consider a thermal quantum harmonic oscillator weakly coupled to a heat bath at a different temperature. We analytically study the quantum heat exchange statistics between the two systems using the quantum-optical master equation. We exactly compute the characteristic function of the heat distribution and show that it verifies the Jarzynski-W\'ojcik fluctuation theorem. We further evaluate the heat probability density in the limit of long thermalization times, both in the low and high temperature regimes, and investigate its time evolution by calculating its first two cumulants.
\end{abstract}

\maketitle

Heat and work are two fundamental  quantities in thermodynamics. While these variables are deterministic in macroscopic systems \cite{cal85}, they become stochastic at the microscopic scale owing to the presence of thermal \cite{sei12,sek98} or quantum \cite{esp09,cam11} fluctuations. A central issue is then to determine their probability distributions. The nonequilibrium work statistics of classical driven  systems has been extensively studied both theoretically and experimentally \cite{jar11,cil13a,cil17}. On the other hand, the investigation of heat fluctuations is more involved, even for simple systems at equilibrium \cite{zon04,zon04a,fog09,cha10,gom11,cil13}. The main reason is that heat depends nonlinearly on position even for a linear system like the harmonic oscillator. The heat distribution has been theoretically and experimentally analyzed for a classical harmonic oscillator  in the overdamped limit in Ref.~\cite{imp07} and in the underdamped regime in Ref.~\cite{mar15}.

At the quantum level, attention has  so far  mostly focused on nonequilibrium work. The   work distributions of  driven quantum oscillators have for instance been theoretically obtained in Refs.~\cite{def08,tal09,def10} and experimentally studied using a trapped ion \cite{an15}. At the same time, the quantum work statistics of a driven two-level system  has been computed  in Refs.~\cite{sol13,hek13} and determined experimentally in  NMR  \cite{bat14} and cold-atom \cite{cer17} setups. Recently, the quantum heat exchange statistics has been examined theoretically for exactly solvable two-level models \cite{gas14,pon15}  and the experimental reconstruction of such a heat distribution  has been reported \cite{pet18}. However, to our knowledge, the heat distribution of  a quantum  harmonic oscillator has neither been calculated  nor measured, despite its essential role in many applications \cite{wei08}.

The aim of this paper is to analytically compute and analyze the properties  of the heat distribution of a thermal quantum harmonic oscillator weakly coupled to a reservoir at a different temperature. To that end, we employ master equation methods of quantum optics \cite{wal08}. We  first determine the exact characteristic function of the heat statistics and demonstrate that it obeys the fluctuation theorem of heat exchange of Jarzynski and Wojcik \cite{jar04}. We additionally derive closed form expressions for the heat distribution in the limit of long interaction times, both in the high and low temperature regimes. We finally study  the time evolution of the  heat probability density by analytically evaluating its first two cumulants.

Let us begin by considering a  quantum harmonic oscillator with frequency $\omega$ and inverse temperature $\beta_1$ weakly coupled to a heat bath at a different inverse temperature $\beta_2$. We model the reservoir  as an infinite set of quantum harmonic oscillators, as commonly done in condensed matter physics \cite{wei08} and quantum optics \cite{wal08}. The Hamiltonian of the combined system is $H = H_1  + H_2 + H_{12}$, where
$H_1 = \hbar \omega (  a^{\dagger} a + 1/2 )$ and $H_2 = \sum_j \hbar \omega_j b_j^{\dagger} b_j $ are the respective Hamiltonians of system and bath, and $H_{12}= \hbar \sum_j \kappa_j ( a^\dagger b_j + a b_j^\dagger)$ describes the    interaction  with coupling parameters $\kappa$ \cite{wal08}. Here $a$ and $b_j$ denote the usual ladder operators.
 System and reservoir  are brought into thermal contact at $t=0$ and let to interact for a duration $t$. Since the oscillator-bath coupling is  weak, heat may be identified with the  energy exchanged between the two. The   heat distribution at time $t$ is accordingly \cite{jar04},
\begin{equation}
\label{1}
P(Q,t) = \sum_{n,m} \delta \left[ Q - (E_m - E_n) \right]  P_{n,m}^t  P_n^0,
\end{equation}
where  $P_n^0 = \exp(-\beta_1 E_n)/Z$ is the initial thermal occupation probability of the oscillator with partition function $Z$, and $P_{m,n}^t$ are the transition probabilities between initial and final states $n$ and $m$ with corresponding energy eigenvalues $E_l=\hbar \omega(l+1/2)$, $l=(n,m)$. The transition probabilities can be explicitly written in terms of the time evolution operator  as $P_{n,m}^t = | \bra{m} U(t) \ket{n} |^2= \bra{m} \rho(t) \ket{m}$, with the density operator $\rho(t) =U(t) \ket{n} \bra{n} U^{\dagger}(t)$. 
We therefore need to determine the diagonal matrix element of the density operator in order to evaluate the heat statistics via Eq.~\eqref{1}.

The   time evolution  of the density operator $\rho$ of a damped  harmonic oscillator in the weak-coupling limit is governed by the quantum-optical master equation \cite{wal08},
\bea
\label{2}
	- \frac{d \rho(t)}{d t} &=& i \omega  \left[a^{\dagger} a, \rho \right] + \frac{\gamma}{2}  \bar n_2 (a a^{\dagger} \rho 
								+ \rho a a^{\dagger} - 2 a^{\dagger} \rho a) 	\nonumber		\\
								&&+  \frac{\gamma }{2} (\bar n_2 + 1) (a^{\dagger} a \rho 
								+ \rho a^{\dagger} a - 2 a \rho a^{\dagger}),
\eea
where  $\bar n_2= [\exp (\beta_2 \hbar \omega) - 1]^{-1}$ is the thermal occupation number at inverse temperature $\beta_2$ and $\gamma$  the damping constant. The quantum master equation \eqref{2} may be solved exactly using generating function techniques \cite{arn96}. Writing concretely the diagonal matrix elements in the form $\bra{m} \rho (\tau) \ket{m} = \sum_{n} X_{m,n}(\tau) p_n(0)$ with $\tau = \gamma t$ and arbitrary initial condition $p_n(0)$, one finds  \cite{arn96},
\begin{align}
	X_{m,n} &= \frac{u^m}{(1+u)^{m+1}} \left( \frac{1 + v}{1 + u}\right)^n  \sum_j \frac{(m+n - j)!}{(n-j)! j! (m-j)! } \nonumber \\
				&	\times \left[- \frac{v(1+u)}{u(1+v)} \right]^j, 
\end{align}
with the two parameters $u$ and $v$ defined as,
\be
\label{4}
u  = \bar n_2 (1 - e^{- \tau}) \quad \text{and}\quad
	v  = \bar n_2 - (\bar n_2 + 1) e^{- \tau}. 
\ee
\begin{figure}[t]
\centering
\includegraphics[width=0.49\textwidth]{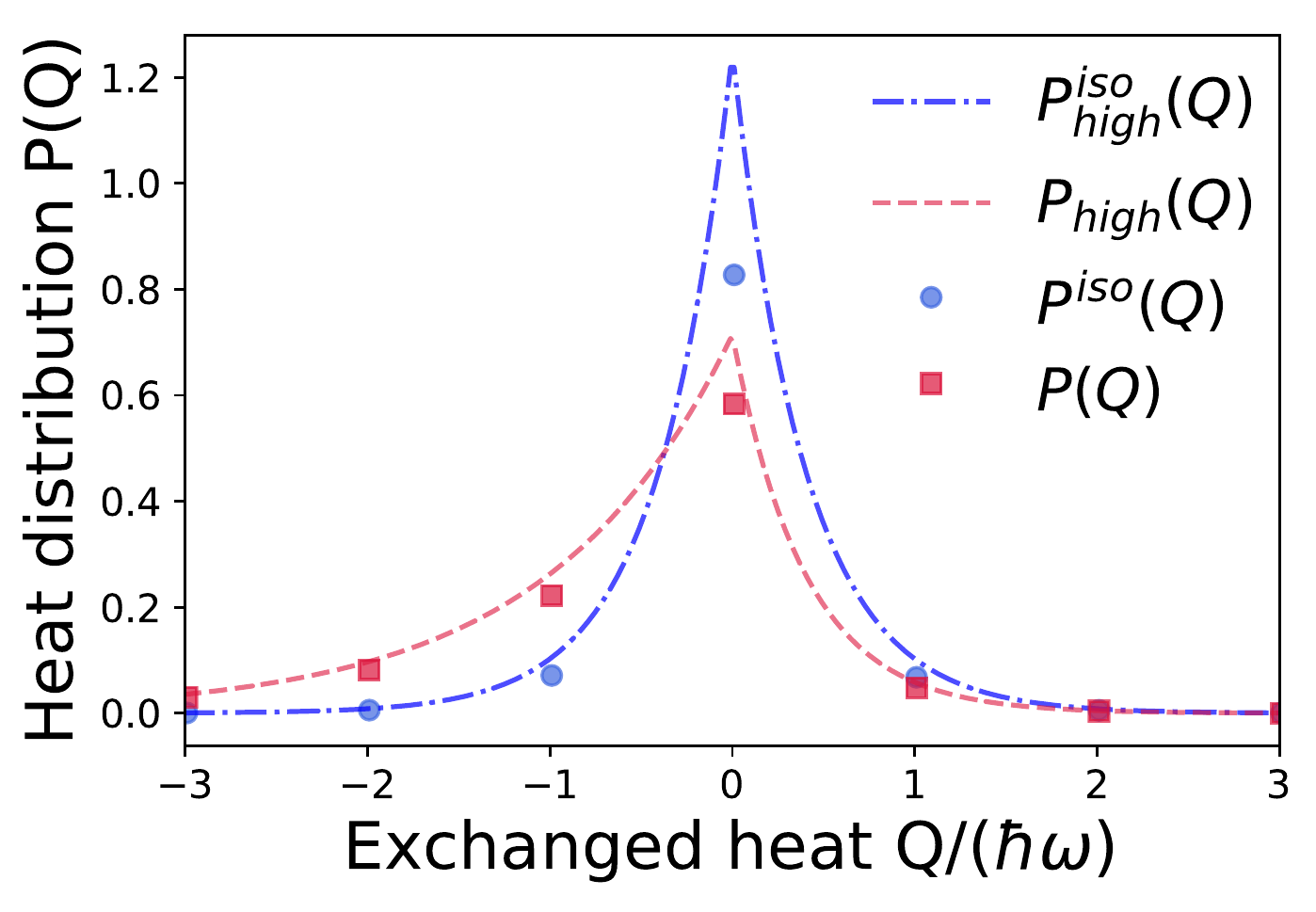}
\caption{Asymptotic quantum quantum heat distribution $P(Q)$, Eq.~\eqref{9}, for a harmonic oscillator at inverse temperature $\beta_1$ weakly coupled to a bath at inverse temperature $\beta_2$ (red squares), compared with the symmetric isothermal heat distribution $P^\text{iso}(Q)$, Eq.~\eqref{11}, obtained for $\beta=\beta_1=\beta_2$ (blue dots). The respective blue dotted-dashed and red dashed lines represent the corresponding classical heat distribution given by Eq.~(12). Parameters are $\beta_1= 1$, $\beta_2 =2.5 $ and $\beta=2.5$.}\label{fig:p_q}
  \end{figure}
The heat distribution \eqref{1} follows with $p_n(0) =1$ as,
\begin{equation}
\label{5}
 	P(Q, \tau) = \frac{1}{Z} \sum_{n,m} \delta [Q - (E_m - E_n)] X_{m,n}(\tau)  e^{- \beta_1 \hbar \omega n }.
\end{equation}
In order to analyze Eq.~\eqref{5}, we introduce the characteristic function $G(\mu, \tau)= \int dQ \exp({i\mu Q) P(Q,\tau)}$ and obtain,
\be
\label{6}
	G(\mu, \tau) 	=\frac{1}{Z}\sum_{n,m} X_{m,n}(\tau)  e^{- \beta_1 \hbar \omega n } e^{i \mu \hbar \omega (m -n) }.
\ee
The three sums appearing in Eq.~\eqref{6} can be performed explicitly, see details below, leading to,
\begin{equation}
\label{7}
	G(\mu,\tau) = \frac{\left(e^{\beta_1 \hbar \omega }-1\right) e^{i \hbar \mu  \omega }}{e^{i \hbar \mu  \omega } [(u+1) e^{\beta_1 \hbar \omega } -u e^{\hbar \omega  (\beta_1+i \mu )}+v]-v-1}
\end{equation}
The above expressions are exact and fully characterize the quantum heat fluctuations of a damped harmonic  oscillator coupled to a reservoir  at a different temperature. The characteristic function \eqref{7} satisfies the symmetry relation $G( i \Delta \beta-\mu,\tau) = G(\mu,\tau)$.  We thus recover the  fluctuation theorem for heat exchange, $P(Q,\tau) /P(-Q,\tau) =  \exp(-\Delta \beta Q) $,  derived by Jarzynski and W\'ojcik \cite{jar04}. In order to gain  additional physical insight about the quantum heat statistics, we will now study different limits where closed form  formulas can be derived.

We start by examining the long-time behavior of the heat statistics.
In the  limit $\tau \rightarrow \infty$, Eq.~\eqref{7} reduces  to,
\bea\label{8}
G(\mu)  &=&  \frac{1}{Z}  \sum_{n,m} e^{\beta_1 \hbar \omega}
	e^{-i \mu \hbar \omega n} \left( \frac{\bar n}{1+\bar n} \right)^m \frac{1}{1 + \bar n}  \\
&=&
 \frac{1 - e^{- \hbar \omega \beta_1 } - e^{- \hbar \omega\beta_2 } + e^{- \hbar \omega (\beta_1 + \beta_2)} }{1 - e^{-\hbar \omega (\beta_2 - i \mu)} - e^{-\hbar \omega (\beta_1 + i \mu)} + e^{-\hbar \omega (\beta_1 + \beta_2)} }.\nonumber
\eea
Taking the inverse Fourier transform, we arrive at the asymptotic quantum heat distribution,
\bea \label{9}
	P(Q) & =& \frac{1 - e^{-\hbar \omega \beta_1} - e^{-\hbar \omega \beta_2} + e^{-\hbar \omega (\beta_1 + \beta_2)}}{1 - e^{-\hbar \omega (\beta_1 + \beta_2)} }\\ 	&\times& \sum_n  \delta (Q - n \hbar \omega) + \delta (Q + n \hbar \omega) 
	 \begin{cases}
		 e^{- \beta_2 Q}, &  Q \geq 0\\ 
		 e^{\beta_1 Q}, &  Q < 0
	 \end{cases}\nonumber
\eea
In the isothermal case, $\beta=\beta_1=\beta_2$, the characteristic function \eqref{8} further simplifies to,
\begin{equation}
\label{10}
	G^\text{iso}(\mu) = \frac{\cosh(\hbar \omega \beta) - 1}{\cosh(\hbar \omega \beta) - \cos(\hbar \omega \mu)}.
\end{equation}
The corresponding   probability distribution then reads,
\bea\label{11}
	P^\text{iso}(Q) &= &\frac{\cosh(\hbar \omega \beta) - 1}{\sinh(\hbar \omega \beta)} e^{- \beta |Q|}\nonumber \\
	&\times& \sum_n \delta (Q - n \hbar \omega) + \delta (Q + n \hbar \omega).
\eea
Equations \eqref{9} and \eqref{11} are shown in Fig.~1. We observe that the two heat distributions are discrete with spacing $\hbar \omega$, as expected for a quantized harmonic oscillator. We further note that they both decay exponentially for positive and negative arguments. In addition, the heat probability density is in general asymmetric, implying a non-zero mean heat current between oscillator and bath, except in the  isothermal  case since no average energy flows between two  objects at the same temperature. 
  
 In the high-temperature limit, $\hbar \omega \beta_{1,2} \ll1$, the discrete heat distribution \eqref{9} becomes continuous and we recover the known classical expression \cite{mar15} by Taylor expanding the exponential functions to lowest order, 
\be
P_\text{high}(Q)=	 \frac{\beta_1\beta_2}{\beta_1+\beta_2}
\begin{cases}
	e^{- \beta_2 Q}, &  Q \geq 0,\\ 
		  e^{\beta_1 Q}, &  Q < 0.
	 \end{cases}
	 \ee
As seen in Fig.~1, the envelops of the classical and quantum heat distributions are similar in shape, in contrast to the work distribution \cite{def08}. The notable difference is that the quantum density is always narrower than the corresponding classical density, owing to the bosonic nature of the harmonic oscillator.
In the opposite low-temperature regime, $\hbar \omega \beta_{1,2} \gg1$, only the  first three delta peaks at $Q=0,\pm \hbar \omega$ contribute significantly to the heat distribution. As a result, we obtain the  heat probability density,
\bea
\label{12}
 	P_\text{low}(Q) &=&  \frac{ \delta(Q) + \delta (Q - \hbar \omega) + \delta (Q + \hbar \omega)}{1 + e^{- \hbar \omega \beta_1} + e^{- \hbar \omega \beta_2}} \nonumber \\
	 &\times&\begin{cases}
		 e^{- \beta_2 Q}, &  Q \geq 0\\ 
		 e^{\beta_1 Q}, &  Q < 0
	 \end{cases}
\eea
Expression \eqref{12} shows that quantum heat is strictly negative when the harmonic oscillator is initially in its ground state. This corresponds to the limiting situation where the quantum oscillator can only absorb energy.

  \begin{figure}[t]
		  \includegraphics[width=0.49\textwidth, clip, trim=0cm 0cm 0cm 0cm]{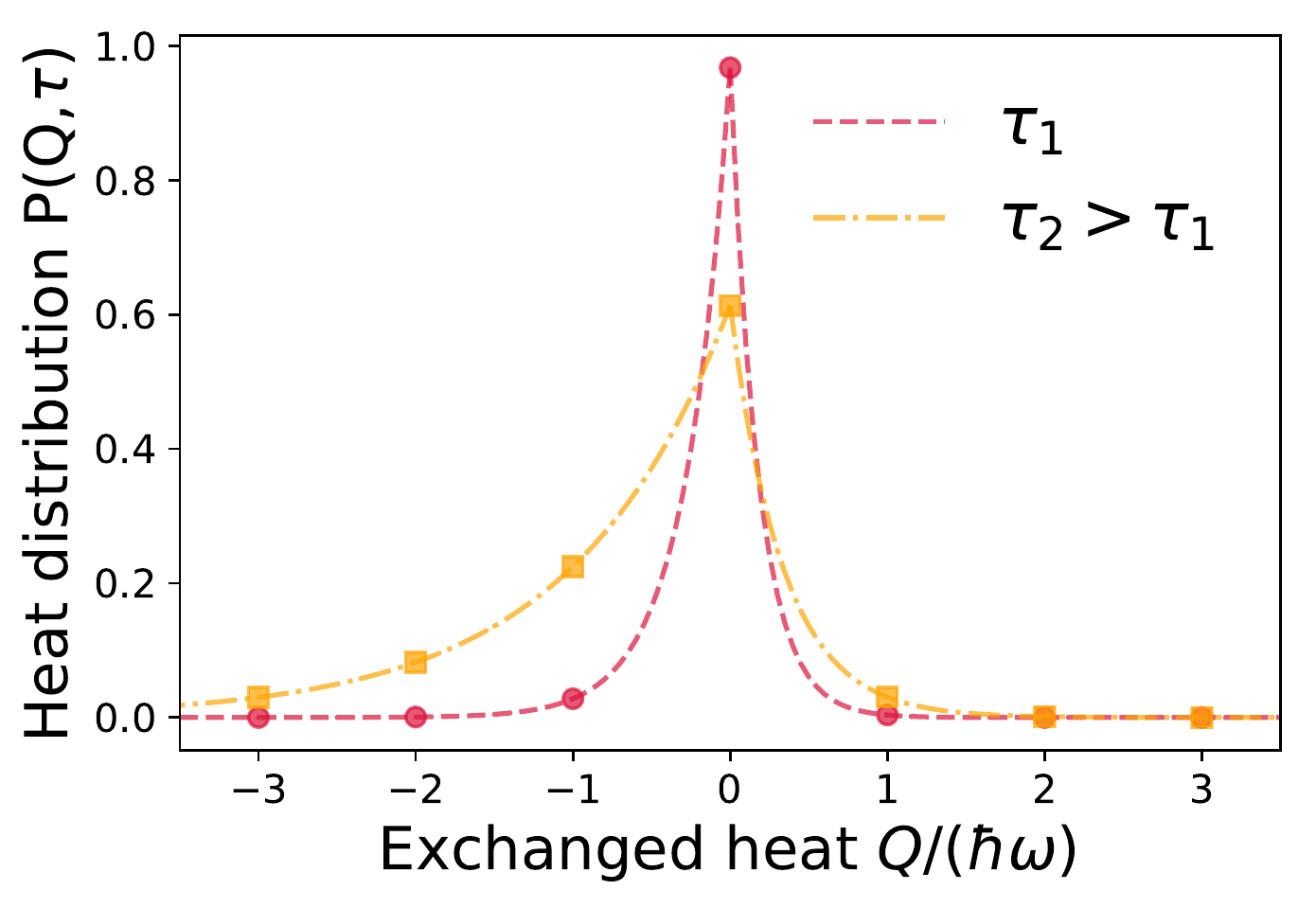}

	\caption{ Evolution of the quantum heat distribution $P(Q,\tau)$ computed as the inverse Fourier transform of the characteristic function $G(\mu,\tau)$, Eq.~\eqref{7}, for two thermalization times $\tau_1= 0.1$ and $\tau_2=2$. Parameters are $\beta_1= 1$, $\beta_2=3$.}
  \end{figure}
  
It does not seem  possible to analytically determine the quantum heat distribution $P(Q,\tau)$ for arbitrary thermalization times $\tau$ (see Fig.~2). In order to study its time evolution, we next   compute its first two cumulants using the formula $\la Q^n\ra(\tau) = i^{-n}  d^{n} G(\mu,\tau)/d \mu^n {|\mu=0}$ \cite{ris89}. We obtain the average heat,
\begin{equation}
\label{13}
		\langle Q \rangle(\tau) = \frac{\hbar \omega  \left(u e^{\beta_1 \hbar \omega }-v-1\right)}{e^{\beta_1 \hbar \omega }-1},
	\end{equation}
and the variance,
\begin{widetext}
\be
\label{14}
	\sigma^2_Q(\tau) = \la Q^2\ra(\tau) - \la Q\ra^2(\tau)  
	=\frac{\hbar^2 \omega ^2 \left[u (u+1) e^{2 \beta_1 \hbar \omega }+(1-u (2 v+3)+v) e^{\beta_1 \hbar \omega }+v^2+v\right]}{\left(e^{\beta_1 \hbar \omega }-1\right)^2}.\ee
\end{widetext}
in terms of the time-dependent parameters $u$ and $v$ given in Eq.~\eqref{4}. The variance increases as a function of time (see Fig.~3), indicating that  the heat distribution widens. This can be physically understood by noting that no heat is exchanged between oscillator and reservoir when they are initially brought into thermal contact. The initially heat distribution is accordingly a Dirac delta with vanishing variance. As time increases, both mean and variance  approach their stationary values exponentially, as expected for a linear system.  The asymptotic long-time limits of Eqs.~\eqref{13} and \eqref{14} are respectively,
\begin{equation}
\label{16}
	\langle Q \rangle = \frac{1}{2} \hbar \omega \left[\coth \left(\frac{\beta_2 \hbar \omega}{2}\right)-\coth \left(\frac{\beta_1 \hbar \omega}{2}\right)\right],
\end{equation}
and
\begin{widetext}
\begin{equation}
\label{17}
	\sigma^2_Q = \frac{\hbar^2 \omega ^2 \left[-4 e^{\hbar \omega  (\beta_1+\beta_2)}+e^{\hbar \omega  (2 \beta_1+\beta_2)}+e^{\hbar \omega  (\beta_1+2 \beta_2)}+e^{\beta_1 \hbar \omega }+e^{\beta_2 \hbar \omega }\right]}{\left(e^{\beta_1 \hbar \omega }-1\right)^2 \left(e^{\beta_2 \hbar \omega }-1\right)^2}.
\end{equation}
\end{widetext}
Equation \eqref{16} is simply the difference between the mean energies  at temperatures $T_2$ and $T_1$ and can be rewritten in terms of the thermal occupation probabilities as $\langle Q \rangle = \hbar \omega (\bar n_2-\bar n_1)$. We additionally notice that the  heat fluctuations, as characterized by the variance, are left invariant when the temperatures of the harmonic oscillator and of the heat reservoir are switched. This is not the case for  the average value of the heat which  changes its sign, indicating a reversal of the energy current.

  \begin{figure}[t]
		  \includegraphics[width=0.49\textwidth, clip, trim=0cm 0cm 0cm 0cm]{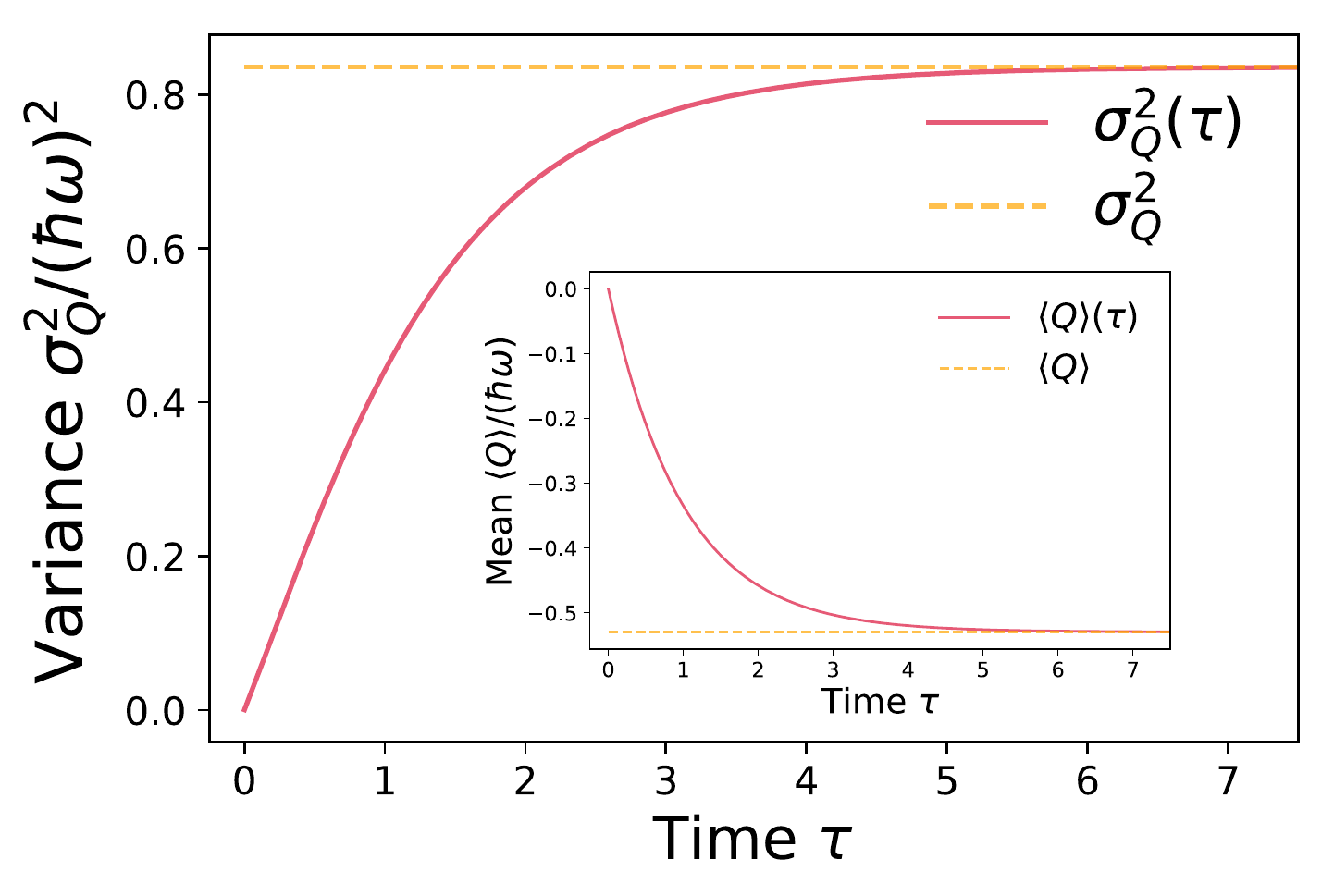}

	\caption{The variance 	$\sigma^2_Q(\tau)$, Eq.~\eqref{14}, (red solid) approaches its steady state value  $\sigma^2_Q$, Eq.~\eqref{17}, (orange dashed) exponentially in time. The inset shows the exponential relaxation of the mean  $\langle Q \rangle(\tau)$, Eq.~\eqref{13}, (red solid) to its asymptotic value $\langle Q \rangle$, Eq.~\eqref{16}, (orange dashed). Same parameters as in Fig.~2.}  \end{figure}

  \textit{Conclusions}.
  We have analytically computed  the characteristic function of the quantum heat statistics of a harmonic oscillator weakly coupled to a heat reservoir at a different temperature. We have first shown that it satisfies the fluctuation theorem of Jarzynski and W\'ojcik \cite{jar04}. We have additionally obtained closed form expressions for the quantum heat  distribution in the asymptotic long-time limit, both  in the low and high temperature regimes. The classical and quantum heat  probability densities have the same exponential, and generally asymmetric, dependence on $Q$. The quantum distribution is discrete with spacing corresponding to the  level interval of  the harmonic oscillator. It is moreover narrower than the classical distribution. We have finally investigated the time evolution of the quantum heat distribution by evaluating its first cumulants. We have shown that the stationary limit is reached exponentially in time.
  
  \textit{Appendix.} Let us sketch the derivation of the characteristic function \eqref{7}. We first write Eq.~\eqref{6} in terms of the ordinary hypergeometric function $F[a,b,c;z]$ \cite{whi27},
\bea
G(\mu,\tau) &=& \frac{1}{Z}\sum_{m,n} \frac{u^m}{(1+u)^{m+1}} \left(\frac{1+v}{1+u}\right)^n e^{-\hbar \omega \beta_1} \nonumber\\
 &\times& e^{i \hbar \omega \mu (m-n)} F \left[ -n,-m,1;y\right],
\eea
where we have defined the variable $y=(u-v)/u(1+v)$. We next use the identity,
\begin{equation}
	F \left[ -n,-m,1; y \right] = \left( 1 - y\right)^{1+m+n} F \left[ 1+n,1+m,1; y\right]
\end{equation}
together with the explicit series representation of the ordinary  hypergeometric function,
\begin{equation}
	F \left[ 1+n,1+m,1; y \right] = \sum_{k=0}^\infty y^k \binom{n+k}{k} \binom{m+k}{k}.
\end{equation}
We then obtain the characteristic function,
\bea
	G(\mu,\tau) &=& \frac{1-y}{Z(1+u)} \sum_k y^k \nonumber \\
	& \times& \sum_n \left( \frac{(1+v)(1-y)}{1+u} e^{-\hbar \omega (\beta_1+i \mu)}\right)^n \binom{n+k}{k}\nonumber  \\
	& \times& \sum_m \left(\frac{u(1-y)}{1+u}e^{i \mu \hbar \omega}\right)^m \binom{m+k}{k}.
\eea
The two sums over $m$ and $n$  are of the form,
\begin{equation}
	\sum_{j=0}^\infty a^j \binom{j+k}{k} = \left( 1 - a \right)^{-k-1}, \quad  |a| \le 1.
\end{equation}
As a consequence, we find,
\begin{equation}
	G(\mu,\tau) =  \frac{1-y}{Z(1+u)} \sum_k y^k (1-B)^{-k-1} (1-C)^{-k-1},
\end{equation}
where we used introduced the two parameters,
\begin{equation}
	B = \frac{u(1-y)}{1+u}e^{i\hbar \omega \mu}, \quad C= \frac{(1+v)(1-y)}{1+u} e^{-\hbar \omega (\beta_1+i\mu)}
\end{equation}
The final sum  is a geometric series. We thus arrive at,
\begin{equation}
	G(\mu,\tau) =  \frac{1-y}{Z(1+u)} \cdot \frac{1}{(1-C)(1-B)-y}.
\end{equation}
The characteristic function \eqref{7} follows by inserting the values of $B$ and $C$ given in Eq.~(28) into Eq.~(29).
  
 We thank Hans C. Fogedby for attracting our attention to the quantum heat distribution. TD acknowledges financial support from the Volkswagen Foundation under project "Quantum coins and nano sensors".\\
  \hspace{.1cm}

\end{document}